\documentclass[3p]{elsarticle}

\usepackage{amsmath,amsfonts,amssymb, amsthm}

\usepackage{cases}

\usepackage{graphicx}   
\usepackage{graphics,color}
\usepackage{setspace}  
\usepackage{caption}
\usepackage{listings}
\usepackage{tabularx}
\usepackage{xcolor}
\usepackage{indentfirst}
\usepackage{subcaption}

\usepackage{multirow}
\usepackage{overpic}

\usepackage{verbatim}

\usepackage{enumerate}

\usepackage{bm}

\usepackage[font=footnotesize,labelfont=bf]{caption}

\newtheorem{remark}{Remark}[section]

\usepackage[version = 4]{mhchem}

\newcommand\dd{\mathrm{d}}

\newcommand\Vvec{\mathbf{V}}

\newcommand\uvec{\bm{u}}
\newcommand\X{\mathbf{X}}
\newcommand\x{\bm{x}}

\newcommand\qvec{\bm{q}}
\newcommand\Qvec{\bm{Q}}  
\newcommand\Uvec{\bm{U}}

\newcommand\rvec{\bm{r}}

\begin{document}

\title{Micro-Macro Modeling of Polymeric Fluids with Multi-Bead Polymer Chain}

\author{Xuelian Bao\footnote{baoxuelian@scut.edu.cn}}
\address{School of Mathematics, South China University of Technology, Guangzhou, Guangdong 510641, People’s Republic of China}

\author{ Lidong Fang\footnote{fanglidong@sufe.edu.cn}}
\address{School of Mathematics, Shanghai University of Finance and Economics, Shanghai, 200433, People’s Republic of China}

\author{ Huaxiong Huang\footnote{hhuang@yorku.ca}}
\address{Zu Chongzhi Center, Duke Kunshan University, Suzhou, 215316, People’s Republic of China; Guandong Key Lab for Data Science and Technology, BNU-HKBU United International College, Zhuhai, 519087,
People’s Republic of China; Department of Mathematics and Statistics, York University, Toronto, M3J 1P3, Canada}


\author{ Zilong Song \footnote{zilong.song@usu.edu}}
\address{Department of Mathematics and Statistics, Utah State University, Logan, 84322, USA }

\author{ Shixin Xu \footnote{shixin.xu@dukekunshan.edu.cn}}
\address{Zu Chongzhi Center, Duke Kunshan University, Suzhou, 215316, People’s Republic of China}

\begin{abstract}
This work extends the classical dumbbell (two-bead) model of polymer chains to a more detailed multi-bead representation, where each polymer chain consists of $N$ beads connected by $N-1$ springs. We develop a thermodynamically consistent micro-macro model based on the energy variational method to describe the coupled dynamics of polymer configurations and fluid flow. The resulting framework captures complex microscopic behaviors, such as bond stretching and alignment under flow, and links them to macroscopic stress responses. 

\end{abstract}

\maketitle

\section{Model Derivation}

In this section, we derive the multi-bead model, where a polymer chain consists of $N$ beads connected by $N-1$ springs on the microscopic scale.
We start from the macroscopic scale, in a bounded domain $\Omega$, we have the law of mass conservation:
\begin{equation}
    \begin{aligned}
    & \rho\frac{D \uvec}{Dt} = \nabla_{\x}\cdot \bm\sigma = \nabla_{\x}\cdot \bm\sigma_s + \nabla_{\x}\cdot \bm\sigma_p, \\
    & \nabla_{\x} \cdot \uvec = 0, 
    \end{aligned}
\end{equation}
where $\rho$ is the fluid density, $\uvec$ is the fluid velocity, and $\frac{D \uvec}{Dt} = \frac{\partial \uvec}{\partial t} + \uvec \cdot \nabla_{\x} \uvec$ is the material derivative. This is the conservation of momentum with stress tensor $\bm \sigma$, which is induced by the viscosity of fluid $\bm \sigma_s$ and polymer $\bm \sigma_p$. 
We set the Dirichlet boundary condition for $\uvec$, i.e. $\uvec|_{\partial \Omega} = 0$.


For the microscopic scale, the bead position is determined by $\rvec_1$, $\rvec_2$, $\cdots$, $\rvec_N$. And the bead-to-bead vector can be determined by $\qvec_N$, $\qvec_2$, $\cdots$, $\qvec_{N-1}$ with $\qvec_i=\rvec_{i+1}-\rvec_i$ with $i=1, 2, \cdots, N-1$. 
In the configurational space, 
let $n(\qvec_1, \qvec_2, \cdots, \qvec_{N-1}, t)$ be the number density of polymers which have end-to-end vectors $\qvec_1, \qvec_2, \cdots, \qvec_{N-1}$ at time $t$. Assume that $f$ only depends on $\qvec_1, \qvec_2, \cdots, \qvec_{N-1}$. 
Then for any $\x \in \Omega$ in the physical space, we have
\begin{equation}\label{int_n}
     \int_{\mathbb{R}^d} \int_{\mathbb{R}^d}\cdots \int_{\mathbb{R}^d} n(\x, \qvec_1, \qvec_2, \cdots, \qvec_{N-1}, t) \dd \qvec_1 \dd \qvec_2 \cdots \dd \qvec_{N-1} = N_q,
\end{equation}
where $N_q$ is the total number of all polymers. 
Denote $f$ as the probability density in the configurational space, i.e., $f(\x, \qvec_1, \qvec_2, \cdots, \qvec_{N-1}, t) = n(\x, \qvec_1, \qvec_2, \cdots, \qvec_{N-1}, t)/N_q$.

Borrowing the flow map definitions from the energetic variational approach \cite{Giga2017,wang2020jcp,wang2021two}, let $\x(\X, t)$ be the flow
map at the physical space and $\qvec_i(\X, \Qvec_i, t)$ be the flow maps at the configurational space ($i=1, 2, \cdots, N-1$),  where $\X$ and $\Qvec_i$ are Lagrangian coordinates in physical and configurational space respectively. 
Hence, we have
$$
f(\x, \qvec_1, \cdots, \qvec_{N-1}, t) = f(\x(\X, t), \qvec_1(\x(\X, t), \Qvec_1, t), \cdots, \qvec_{N-1}(\x(\X, t), \Qvec_{N-1}, t), t)
$$
with $\x \in \Omega$ and $\x(\X, t=0)=\X$. 

Assume arbitrary unit volume $\Omega_{\Qvec}$ in the Lagrangian coordinate of the configurational space, we have 
\begin{equation}\label{f0}
    \begin{aligned}
       \int \cdots  \int_{\Omega_{\Qvec}} f_0(\X, \Qvec_1, \cdots \Qvec_{N-1}) \dd\Qvec_1 \cdots  \dd\Qvec_{N-1} & = \int \cdots \int_{\Omega_{\qvec}} f(\x, \qvec_1, \cdots \qvec_{N-1}, t) \dd\qvec_1 \cdots \dd\qvec_{N-1}\\
        & = \int \cdots \int_{\Omega_{\Qvec}} f \bigg( \prod_{i=1}^{N-1} det G_i \bigg)  \  \dd\Qvec_1 \cdots \dd\Qvec_{N-1}
    \end{aligned}
\end{equation}
due to mass conservation, where $f_0(\X, \Qvec_1, \cdots, \Qvec_{N-1}) = f(\X, \Qvec_1, \cdots, \Qvec_{N-1}, 0)$  is the initial number distribution function, $\Omega_{\qvec}$ is the corresponding volume of $\Omega_{\Qvec}$ at time $t$ and the Jacobi matrix $G_i = \nabla_{\Qvec_i} \qvec_i$.

Since $\Omega_{\Qvec}$ and $f_0(\X, \Qvec_1, \cdots, \Qvec_{N-1})$ are independent of time $t$, taking derivative of \eqref{f0}with respect to $t$, we have
\begin{equation}\label{dtf0}
    \begin{aligned}
        0 & = \frac{\dd}{\dd t} \int \cdots \int_{\Omega_{\Qvec}} f \bigg( \prod_{i=1}^{N-1} det G_i \bigg)  \  \dd\Qvec_1 \cdots \dd\Qvec_{N-1}\\
        & = \int \cdots \int_{\Omega_{\Qvec}} \frac{d}{dt}\bigg( f  \prod_{i=1}^{N-1} det G_i \bigg) \ \dd\Qvec_1 \cdots \dd\Qvec_{N-1}\\
        & = \int \cdots \int_{\Omega_{\Qvec}}
        \left(f_t +\uvec\cdot \nabla_{\x} f+ \sum_{i=1}^{N-1} \Vvec_i\cdot\nabla_{\qvec_i} f +G_i^{-T}:\frac{dG_i}{dt}f  \right)\prod_{i=1}^{N-1} det G_i\ \dd\Qvec_1 \cdots \dd\Qvec_{N-1}\\
        & = \int \cdots \int_{\Omega_{\Qvec}}
        \left(f_t +\uvec\cdot \nabla_{\x} f+ \sum_{i=1}^{N-1} \Vvec_i\cdot\nabla_{\qvec_i} f +f\nabla_{\qvec_i}\cdot\Vvec_i \right)\prod_{i=1}^{N-1} det G_i\ \dd\Qvec_1 \cdots \dd\Qvec_{N-1}\\
    \end{aligned}
\end{equation}
where $\Vvec_i  = \dot{\qvec_i} = \qvec_{it} + \uvec \cdot \nabla_{\x} \qvec_i$ is the microvelocity.

Denote $\tilde{\Vvec}_i$ be the difference between the microvelocity and the macroscopic induced velocity.
Due to the Cauchy-born rule, $\qvec_i = F \bm Q_i$ and $F$ is the deformation gradient tensor $F = \nabla_{\X} \x$, which indicates that the macroscopic induced velocity equals to $\frac{d}{dt} (F \bm Q_i) = (\frac{d}{dt} F) \bm Q_i =  (\nabla_{\x} \uvec)^T \cdot \qvec_i$. 
Hence, 
\begin{equation}\label{macroinduced}
 \tilde{\Vvec}_i = \Vvec_i - (\nabla_{\x} \uvec)^T \cdot \qvec_i.
\end{equation}
Then due to the arbitrariness of $\Omega_{\Qvec}$, the law of conservation  in the microscale yields  
\begin{equation}\label{conservationf}
  f_t + \nabla_{\x} \cdot ( f\uvec)  + \sum_{i=1}^{N-1} \nabla_{\qvec_i} \cdot ( (\nabla_{\x} \uvec)^T \cdot \qvec_i f) = - \sum_{i=1}^{N-1} \nabla_{\qvec_i}\cdot (f \tilde{\Vvec}_i),  
\end{equation}
where the incompressibility condition $\nabla_{\x} \cdot \uvec = 0$ is used.

We then derive the micro-macro multi-bead model via the energy variational method \cite{shen2020energy,shen2022energy, bao2024prf}.  We first have the energy-dissipation law
\begin{equation}\label{dissipationlaw}
    \frac{\dd F_{total} }{\dd t} = -\Delta.
\end{equation}
The total energetic functional  is defined as the sum of the kinetic energy of the fluid on the macroscale,  the mixing energy and internal energy of the polymer on the microscale\cite{bird1987dynamics},
 
\begin{equation}
\begin{aligned}
& F_{total} =  F_{macro} + \int_{\Omega} F_{micro} \dd \x\\
& \quad = \int_{\Omega} \frac{1}{2} \rho |\uvec|^2 \dd \x + \int_{\Omega} \lambda_p\bigg[ \int_{\mathbb{R}^d} \cdots \int_{\mathbb{R}^d} (k_B T f (\ln (f/f_{\infty}) -1)+\Psi f) \dd \qvec_1 \cdots \dd \qvec_{N-1} \bigg] \ \dd \x,  
\end{aligned}
\end{equation}
where $\Psi = \Psi(\qvec_1, \cdots \qvec_{N-1})$ is the internal energy density functional,  $\lambda_p$ is a parameter related to the polymer density, $f_{\infty}$ is the distribution in the reference state, which is a constant with respect to the configurational space. 
Here we assume that the properties of the polymer chains are independent of the position of the chain on the microscopic scale. They only depend on the relative positions of every two beads. 
 The  chemical potential is defined according to the energetic functional as follows:
$$
\mu_f = \frac{\delta F_{total}}{\delta f} = \lambda_p [k_B T \ln (f/f_\infty) +\Psi].
$$

The dissipative functional $\Delta$ consists of fluid friction on the macroscale, and dissipation induced by the relative friction of microscopic polymers to the macroscopic flow: 
 
\begin{equation}\label{dissipationdef}
\begin{aligned}
    \Delta &= \int_{\Omega} 2\eta_s |\bm D_{\eta}|^2 d\x +    \lambda_p \zeta \int_{\Omega} \int_{\mathbb{R}^d}v\cdots \int_{\mathbb{R}^d}   \sum_{i=1}^N f  |\dot{\rvec}_i - (\nabla_{\x} \uvec)^T \cdot \rvec_i |^2  \ \dd \qvec_1 \cdots \dd \qvec_{N-1} \dd \x\\
    & =  \int_{\Omega} 2 \eta_s |\bm D_{\eta}|^2 \dd \x  +  \lambda_p \zeta  \int_{\Omega}   \int_{\mathbb{R}^d} \cdots  \int_{\mathbb{R}^d} f  \sum_{i=1}^{N-1} \sum_{j=1}^{N-1} 
    C_{ij}( \dot{\qvec}_i - (\nabla_{\x} \uvec)^T \cdot \qvec_i) \cdot ( \dot{\qvec}_j - (\nabla_{\x} \uvec)^T \cdot \qvec_j)  \dd \qvec_1 \cdots \dd \qvec_{N-1} \dd \x  \\
    & =  \int_{\Omega} 2 \eta_s |\bm D_{\eta}|^2 \dd \x  +  \lambda_p \zeta  \int_{\Omega}   \int_{\mathbb{R}^d} \cdots  \int_{\mathbb{R}^d} f  \sum_{i=1}^{N-1} \sum_{j=1}^{N-1} 
    C_{ij} \tilde{\Vvec}_i  \cdot  \tilde{\Vvec}_j   \dd \qvec_1 \cdots \dd \qvec_{N-1} \dd \x, 
    \end{aligned}
\end{equation}
where $\eta_s$ is the viscosity of the fluid, $\bm D_{\eta} = \frac{1}{2} (\nabla_{\x} \uvec + (\nabla_{\x} \uvec)^T)$ is the strain rate, $\lambda_p$ and $\zeta$ are constants 
related to the polymer density and polymer relaxation time, respectively.
%
$ C_{ij}$ are the elements of the $(N-1) \times (N-1)$ Kramers matrix, defined as follows, 
$$
C_{ij} = \left\{
\begin{aligned}
& i(N-j)/N, \qquad \mbox{if} \ i \leq j,\\
& j (N-i)/N, \qquad \mbox{if} \ j \leq i.
\end{aligned}
\right.
$$
And the Kramers matrix is inverse of $(N-1) \times (N-1)$ Rouse matrix $A$, with
$A_{ik}$ the elements of the Rouse matrix defined by
$$
A_{ij} = \left\{
\begin{aligned}
& 2 \qquad \quad \mbox{if} \ i=j,\\
& -1 \qquad \mbox{if} \ i=j \pm 1, \\
& 0 \qquad \quad \mbox{otherwise.} 
\end{aligned}
\right.
$$

\begin{remark}

The critical step in \eqref{dissipationdef} involves transforming the bead position 
 $\rvec_i$ to $\qvec_i$ for the multi-bead model. Precisely, we take 3-bead and 4-bead models as an example here. 
For the 3-bead model, the Kramers and Rouse matrixes can be written as
\begin{equation}
C =
\left[ \begin{array}{ccc}
2/3 & 1/3 \\
1/3& 2/3
\end{array}
\right ], \quad
A =
\left[ \begin{array}{cc}
2 &  -1 \\
-1 & 2\\
\end{array} 
\right ].
\end{equation}
And thus, the dissipation term can be written as
\begin{equation}
\begin{aligned}
    \Delta &= \int_{\Omega} 2\eta_s |\bm D_{\eta}|^2 d\x +   \int_{\Omega} \int_{\mathbb{R}^d}  \int_{\mathbb{R}^d}  \lambda_p \zeta \sum_{i=1}^3 f  |\dot{\rvec}_i - (\nabla_{\x} \uvec)^T \cdot \rvec_i |^2  \ \dd \qvec_1  \dd \qvec_2 \dd \x\\
    & =  \int_{\Omega} 2 \eta_s |\bm D_{\eta}|^2 \dd \x  +\lambda_p \zeta  \int_{\Omega}   \int_{\mathbb{R}^d}  \int_{\mathbb{R}^d} \frac{2}3 f \bigg( | \dot{\qvec}_1 - (\nabla_{\x} \uvec)^T \cdot \qvec_1|^2 + | \dot{\qvec}_2 - (\nabla_{\x} \uvec)^T \cdot \qvec_2|^2 \\
    & \qquad +   ( \dot{\qvec}_1 - (\nabla_{\x} \uvec)^T \cdot \qvec_1)  \cdot ( \dot{\qvec}_2 - (\nabla_{\x} \uvec)^T \cdot \qvec_2) \bigg) \dd \qvec_1  \dd \qvec_2  \dd \x .
    \end{aligned}
\end{equation}

Similarly, for the 4-bead model, the Kramers and Rouse matrixes are
\begin{equation}
C =
\left[ \begin{array}{ccc}
3/4 & 1/2 &1/4  \\
1/2 &1 & 1/2\\
1/4 &1/2 &3/4 
\end{array}
\right ], \quad
A =
\left[ \begin{array}{ccc}
2 &  -1 & 0\\
-1 & 2 & -1\\
0 & -1 & 2
\end{array} 
\right ], 
\end{equation}
and the dissipation term
\begin{equation}
\begin{aligned}
    \Delta &= \int_{\Omega} 2\eta_s |\bm D_{\eta}|^2 d\x +   \int_{\Omega} \int_{\mathbb{R}^d}  \int_{\mathbb{R}^d}  \int_{\mathbb{R}^d} \lambda_p \zeta \sum_{i=1}^4 f  |\dot{\rvec}_i - (\nabla_{\x} \uvec)^T \cdot \rvec_i |^2  \ \dd \qvec_1  \dd \qvec_2 \dd \qvec_3 \dd \x\\
    & =  \int_{\Omega} 2 \eta_s |\bm D_{\eta}|^2 \dd \x  + \lambda_p \zeta  \int_{\Omega}   \int_{\mathbb{R}^d}  \int_{\mathbb{R}^d}  \int_{\mathbb{R}^d} \frac{1}4 f \bigg( 3 | \dot{\qvec}_1 - (\nabla_{\x} \uvec)^T \cdot \qvec_1|^2 + 4| \dot{\qvec}_2 - (\nabla_{\x} \uvec)^T \cdot \qvec_2|^2  + 3| \dot{\qvec}_3 - (\nabla_{\x} \uvec)^T \cdot \qvec_3|^2 \\
    & \qquad +   4( \dot{\qvec}_1 - (\nabla_{\x} \uvec)^T \cdot \qvec_1)  \cdot ( \dot{\qvec}_2 - (\nabla_{\x} \uvec)^T \cdot \qvec_2)  + 2( \dot{\qvec}_1 - (\nabla_{\x} \uvec)^T \cdot \qvec_1)  \cdot ( \dot{\qvec}_3 - (\nabla_{\x} \uvec)^T \cdot \qvec_3) \\
    & \qquad +  4 ( \dot{\qvec}_2 - (\nabla_{\x} \uvec)^T \cdot \qvec_2)  \cdot ( \dot{\qvec}_3 - (\nabla_{\x} \uvec)^T \cdot \qvec_3) \bigg)    
     \dd \qvec_1  \dd \qvec_2   \dd \qvec_3 \dd \x.  
    \end{aligned}
\end{equation}

\end{remark}

With definitions of the total energy functional and the dissipation term, now we are ready to derive the micro-macro multi-bead model via the energy variational method. The definition of the total energy functional yields 
\begin{equation}\label{dEdt}
    \frac{\dd}{\dd t} F_{total} = I_1+  I_2.
\end{equation}
For the first term 
\begin{equation}\label{I1}
\begin{aligned}
    I_1 &= \frac{\dd}{\dd t}F_{macro} = \int_{\Omega} \rho \uvec \cdot \uvec_t \mathrm{d}\x \\
    & = \int_{\Omega} \uvec\cdot(-\rho \uvec \cdot \nabla_{\x} \uvec + \nabla_{\x} \cdot \bm \tau_s + \nabla_{\x} \cdot \bm \tau_p) \mathrm{d}\x\\
    & = \int_{\Omega} p(\nabla_{\x} \cdot \uvec) d\x - \int_{\Omega} \nabla_{\x} \uvec :(\bm \tau_s + \bm \tau_p) \mathrm{d}\x,
\end{aligned}
\end{equation}
where the pressure $p$ is induced as a Lagrange multiplier for incompressibility.
The second term is calculated with the law of conservation on the microscale \eqref{conservationf}:
 \begin{equation*}
    \begin{aligned}
    I_2
    & = \frac{\dd}{\dd t}\int_{\Omega}  F_{micro} \dd \x\\ 
     & = \int_{\Omega} \int_{\mathbb{R}^d} \cdots \int_{\mathbb{R}^d} \mu_f \bigg(-\nabla_{\x} \cdot (\uvec f) -  \sum_{i=1}^{N-1} \nabla_{\qvec_i}\cdot( (\nabla_{\x} \uvec)^T \cdot \qvec_i f)
     - \sum_{i=1}^{N-1} \nabla_{\qvec_i}\cdot(f \tilde{\Vvec}_i)\bigg) \dd \qvec_1 \cdots \dd \qvec_{N-1} \dd \x \\ 
    &= \lambda_p\int_{\Omega} \int_{\mathbb{R}^d} \cdots \int_{\mathbb{R}^d}  k_B T\nabla_{\x} f\cdot\uvec \dd \qvec_1 \cdots \dd \qvec_{N-1} \dd \x +\int_{\Omega} \int_{\mathbb{R}^d} \cdots\int_{\mathbb{R}^d} \bigg( \sum_{i=1}^{N-1} \lambda_p(k_B T\nabla_{\qvec_i} f+f\nabla_{\qvec_i}\Psi)\\
    & \qquad \cdot((\nabla_{\x} \uvec)^T \cdot \qvec_i) +f \nabla_{\qvec_i}\mu_f\cdot \tilde{\Vvec}_i \bigg) \dd \qvec_1 \cdots \dd \qvec_{N-1} \dd \x\\
    &= \int_{\Omega} \int_{\mathbb{R}^d} \cdots \int_{\mathbb{R}^d}  \bigg(\underbrace{\sum_{i=1}^{N-1}  \lambda_p(k_B T\nabla_{\qvec_i} f+f\nabla_{\qvec_i}\Psi)\cdot( (\nabla_{\x} )\uvec)^T  \cdot \qvec_i}_{H_1} +\underbrace{f \nabla_{\qvec_i}\mu_f\cdot \tilde{\Vvec}_i}_{H_2}\bigg)\dd\qvec_1 \cdots \dd\qvec_{N-1} \dd\x, 
    \end{aligned}
\end{equation*}
where the incompressibility $\nabla_{\x} \cdot \uvec =0$ and $\Psi = \Psi(\qvec_1, \cdots, \qvec_{N-1})$ are used.

For the first term in $I_2$, we have
\begin{eqnarray}
H_1 &=& \int_{\Omega} \lambda_p  \int_{\mathbb{R}^d} \cdots\int_{\mathbb{R}^d} (  \sum_{i=1}^{N-1} k_B T  \nabla_{\qvec_i} f + f \nabla_{\qvec_i}\Psi) \cdot ((\nabla_{\x} \uvec)^T \cdot \qvec_i) \dd\qvec_1 \cdots \dd\qvec_{N-1} \dd \x\nonumber \\
&=& \lambda_p\int_{\Omega} \int_{\mathbb{R}^d} \cdots \int_{\mathbb{R}^d} \sum_{i=1}^{N-1} -k_B T f \nabla_{\qvec_i} \cdot ((\nabla_{\x} \uvec)^T \cdot \qvec_i)
\dd \qvec_1 \cdots \dd \qvec_{N-1} \dd \x\nonumber\\
&& \quad + \lambda_p \int_{\Omega} \int_{\mathbb{R}^d} \cdots \int_{\mathbb{R}^d} \sum_{i=1}^{N-1} \nabla_{\x} \uvec :[\nabla_{\qvec_i} \Psi \otimes \qvec_i ]f  \dd\qvec_1 \cdots \dd\qvec_{N-1} \dd\x\nonumber\\
& =&\sum_{i=1}^{N-1} \lambda_p \int_{\Omega} \int_{\mathbb{R}^d} \cdots \int_{\mathbb{R}^d}  \nabla_{\x} \uvec :[\nabla_{\qvec_i} \Psi \otimes \qvec _i]f  \dd \qvec_1 \cdots \dd \qvec_{N-1} \dd\x.\nonumber
\end{eqnarray}
Here, $\otimes$ denotes a tensor product and $\uvec\otimes\mathbf{v}$ is a matrix $(u_i v_j)$ for two vectors $\uvec$ and $\mathbf{v}$.
And here we use the fact that 
$$
\nabla_{\qvec_i} \cdot ( (\nabla_{\x} \uvec)^T \cdot \qvec_i ) = \nabla_{\x} \cdot \uvec = 0.
$$
Thus, we have 
%
\begin{equation}\label{I2} 
I_2  = \int_{\Omega} \int_{\mathbb{R}^d}  \cdots \int_{\mathbb{R}^d}    \sum_{i=1}^{N-1} \lambda_p \nabla_{\x} \uvec :[\nabla_{\qvec_i} \Psi \otimes \qvec_i ]f + f \nabla_{\qvec_i}\mu_f\cdot \tilde{\Vvec}_i
\dd \qvec_1 \cdots \dd \qvec_{N-1} \dd\x.
\end{equation}
Substituting Eqs. \eqref{I1} and \eqref{I2} into Eq. \eqref{dEdt}, we have
 
\begin{equation}
\begin{aligned}
   \frac{\dd}{\dd t} F_{total} &= I_1+   I_2\\
    & = \int_{\Omega} p(\nabla_{\x} \cdot \uvec) \dd\x \\
    & \quad + \int_{\Omega} \nabla_{\x} \uvec :\bigg[   \sum_{i=1}^{N-1} \lambda_p \int_{\mathbb{R}^d} \cdots \int_{\mathbb{R}^d}  [\nabla_{\qvec_i} \Psi \otimes \qvec_i]f  \dd \qvec_1 \cdots \dd \qvec_{N-1} - 
    \bm \tau_s - \bm \tau_p\bigg] \dd\x \\
    & \quad + \int_{\Omega} \int_{\mathbb{R}^d} \cdots \int_{\mathbb{R}^d}   \sum_{i=1}^{N-1} f \nabla_{\qvec_i}\mu_f\cdot \tilde{\Vvec}_i  \dd \qvec_1 \cdots \dd \qvec_{N-1} \dd\x.
\end{aligned}
\end{equation}

Comparing with the energy dissipation law \eqref{dissipationlaw} and the definition of the dissipation term \eqref{dissipationdef}, we obtain
\begin{equation*}\label{jdef}
    \begin{aligned}
    &\bm \tau_s = 2\eta_s \bm D_{\eta}-p\bm I,\\
    &\bm \tau_p = \sum_{i=1}^{N-1} \lambda_p \int_{\mathbb{R}^d} \cdots \int_{\mathbb{R}^d}  [\nabla_{\qvec_i} \Psi \otimes \qvec_i]f  \dd \qvec_1 \cdots \dd \qvec_{n-1},\\
     &\tilde{\Vvec}_i = -\frac{1}{\zeta \lambda_p} \sum_{k=1}^{N-1} A_{ik}  \nabla_{\qvec_k} \mu_f,\\
     & \nabla_{\qvec_i} \mu_f =-\zeta \lambda_p  \sum_{j=1}^{N-1} C_{ij} \tilde{\Vvec}_j.
    \end{aligned}
\end{equation*}



Then the micro-macro multi-bead model is given by \cite{bird1987}
\begin{equation}\label{CPsystemwithf_1}
\left\{
\begin{aligned}
&\rho(\uvec_t+\uvec \cdot \nabla_{\x} \uvec)+\nabla_{\x} p= \eta_s \Delta_{\x} \uvec+\nabla_{\x} \cdot {\bm \tau}_p,\\
& \nabla_{\x}\cdot\uvec=0,\\
&{\bm \tau}_p = \sum_{i=1}^{N-1}  \lambda_p \mathbb{E} (\nabla_{\qvec_i} \Psi \otimes \qvec_i )  = \sum_{i=1}^{N-1} \lambda_p \int_{\mathbb{R}^d} \cdots \int_{\mathbb{R}^d} f \nabla_{\qvec_i} \Psi\otimes \qvec_i \dd \qvec_1 \cdots \dd \qvec_{N-1},\\
&f_t + \nabla_{\x} f \cdot \uvec + \sum_{i=1}^{N-1} \nabla_{\qvec_i} \cdot ( (\nabla_{\x} \uvec)^T \cdot  \qvec_i f) =  \frac{1}{\zeta}  \sum_{i=1}^{N-1} \nabla_{\qvec_i} \cdot \bigg(  \sum_{k=1}^{N-1}   A_{ik}  [k_B T \nabla_{\qvec_k} f+ f \nabla_{\qvec_k} \Psi] \bigg).
\end{aligned}
\right.
\end{equation}

\begin{remark}
    The derived micro-macro multi-bead model \eqref{CPsystemwithf_1} for polymeric fluids based on the energy variational method is the same as that derived from the kinetic theory \cite{bird1987}. 
    And the derived model  \eqref{CPsystemwithf_1} reduces to the simple two-bead model mentioned in Ref. \cite{wang2020jcp, wang2021two,bao2025jcp} when $N=2$, since the Rouse matrix is $A_{11}=2$. 
\end{remark}

\section{ Non-dimensionalization}

In this section, we derive the dimensionless form of the micro-macro system given in Eq. \eqref{CPsystemwithf_1}. To this end, we start from the dimensionless formulation of the energy-dissipation law and introducing the corresponding dimensionless variables \cite{le2012micro}:
\begin{equation}\label{dimensionlessvariable}
\tilde{\x} = \frac{\x}{L}, \quad \tilde{\uvec} = \frac{\uvec}{U}, \quad \tilde{t} = \frac{t}{T}, \quad T = \frac{L}{U},
\end{equation}
where $L$, $T$, $U$ are the characteristic length, time and velocity. Then, we define
$\lambda = \zeta/4H$, $\eta_p = \lambda_p k_B T \lambda$ and $\eta = \eta_s + \eta_p$, where $\eta_p$ is the parameter related to the polymer viscosity and $\eta$ the total viscosity.
And in order to distinguish from different scales, we introduce the following dimensionless variable in the microscopic space: 
$$
\tilde{\qvec}_i=\frac{\qvec_1}{\tilde{L}},  \qquad i=1, 2, \cdots, N-1.
$$

Then the total energy can be written as 
\begin{equation}
\begin{aligned}
&  F_{total} =  F_{macro} + \int_{\Omega} F_{micro} \dd \x\\
& \quad = \int_{\Omega} \frac{1}{2} \rho |\uvec|^2 + \lambda_p\bigg[ \int_{\mathbb{R}^d} \cdots \int_{\mathbb{R}^d} (k_B T f (\ln (f/f_{\infty}) -1)+\Psi f) \dd \qvec_1 \cdots \dd \qvec_{N-1} \bigg] \ \dd \x,  
\end{aligned}
\end{equation}
Denote
$$
\begin{aligned}
\tilde{F} &=   \int_{\mathbb{R}^d} \cdots \int_{\mathbb{R}^d} (k_B T f (\ln (f/f_{\infty}) -1)+\Psi f) \dd \qvec_1 \cdots \dd \qvec_{N-1} \\
&= \int_{\mathbb{R}^d} \cdots \int_{\mathbb{R}^d} (k_B T f (\ln (f/f_{\infty}) -1)+\Psi f) (\tilde{L}^d)^{N-1} \dd \tilde{\qvec}_1 \cdots \dd \tilde{\qvec}_{N-1}.
\end{aligned}
$$
Notice that 
$$
 \int_{\mathbb{R}^d} \cdots \int_{\mathbb{R}^d}  f (\x, \qvec_1, \cdots, \qvec_{N-1}, t) \dd \qvec_1 \cdots \dd \qvec_{N-1} = 1,
$$
namely, 
$$
 \int_{\mathbb{R}^d} \cdots \int_{\mathbb{R}^d}  f(\tilde{L}^d)^{N-1} \dd \tilde{\qvec}_1 \cdots \dd \tilde{\qvec}_{N-1} = 1.
$$
And thus, denote $\tilde{f} = f (\tilde{L}^d)^{N-1}$, we obtain 
$$
\tilde{F}= \int_{\mathbb{R}^d} \cdots \int_{\mathbb{R}^d} k_B T \tilde{f} (\ln ( \tilde{f}/ \tilde{f}_{\infty}) -1)+\Psi  \tilde{f}   \dd \tilde{\qvec}_1 \cdots \dd \tilde{\qvec}_{N-1}.
$$

Since for the Hookean potential, 
$$
\Psi = \frac{1}2 H \sum_{i=1}^N |\qvec_i|^2 = \frac{1}2 H  \tilde{L}^2 \sum_{i=1}^N |\tilde{\qvec}_i|^2 =H \tilde{L}^2 \tilde{\Psi}, 
$$
similarly, for the potential $\Psi$, we can obtain 
$$
\Psi = H \tilde{L}^{N-1} \tilde{\Psi}. 
$$
%
Here, the parameter $\alpha_1$ is introduced, s.t. $H \tilde{L}^2 = \alpha_1 k_B T$, then $\tilde{E}$ can be written as
$$
\tilde{F} = \int_{\mathbb{R}^d} \cdots \int_{\mathbb{R}^d} k_B T \bigg[ \tilde{f} (\ln  \tilde{f}-C) + \alpha_1 \tilde{\Psi} \tilde{f}  \bigg]  \dd \tilde{\qvec}_1 \cdots \dd \tilde{\qvec}_{N-1}. 
$$
And thus, in the three-dimensional physical space, the time derivative of the total energy can be rewritten as
%
$$
\frac{\dd}{\dd t} F_{total} 
= UL^2  \frac{\dd}{\dd \tilde{t}}  \int_{\tilde{\Omega}}    \frac{1}{2} \rho U^2 |\tilde{\uvec}|^2 + \lambda_p k_B T \bigg[ \int_{\mathbb{R}^d} \cdots \int_{\mathbb{R}^d}  \tilde{f} (\ln  \tilde{f}-C) + \alpha_1 \tilde{\Psi} \tilde{f}  \dd \tilde{\qvec}_1 \cdots \dd \tilde{\qvec}_{N-1} \bigg]  \ \dd \tilde{\x}
$$

For the dissipation term, 
%
since $\nabla \uvec  = \frac{U}{L} \tilde{\nabla} \tilde{\uvec}$ and 
$$
(\nabla_{\x} \uvec)^T \cdot \qvec_i = \frac{U}{L} (\tilde{\nabla} \tilde{\uvec})^T \cdot \tilde{L} \tilde{\qvec}_i = \frac{U \tilde{L}}{L} (\tilde{\nabla} \tilde{\uvec})^T \cdot  \tilde{\qvec}_i.
$$
We then introduce another parameter $\alpha_2$ to describe the scale between micro and macro scales, namely, $\tilde{L} = \alpha_2 L$. And thus, we obtain
$$
(\nabla_{\x} \uvec)^T \cdot \qvec_i = \alpha_2 U (\tilde{\nabla} \tilde{\uvec})^T \cdot  \tilde{\qvec}_i
$$
and
$$
\dot{\qvec}_i=\frac{\tilde{L}}{T} \dot{\tilde{\qvec}}_i = \frac{\tilde{L}}{L/U} \dot{\tilde{\qvec}}_i = U \alpha_2 \dot{\tilde{\qvec}}_i.
$$
Therefore, the dissipation term
\begin{equation*}
\begin{aligned}
    \Delta 
    &= \int_{\tilde{\Omega}}  \eta_s U^2 L |\tilde{\nabla} \tilde{\uvec}|^2 + \lambda_p \zeta  (\alpha_2 U)^2 L^3  \int_{\mathbb{R}^d} \cdots \int_{\mathbb{R}^d} \tilde{f}
     \sum_{i=1}^{N-1} \sum_{j=1}^{N-1} C_{ij} ( \dot{\tilde{\qvec}}_i - (\tilde{\nabla} \tilde{\uvec})^T \cdot  \tilde{\qvec}_i) \\
     & \cdot ( \dot{\tilde{\qvec}}_j - (\tilde{\nabla} \tilde{\uvec})^T \cdot  \tilde{\qvec}_j ) \dd \tilde{\qvec}_1 \cdots \dd \tilde{\qvec}_{N-1} \dd \tilde{\x}. \\
    \end{aligned}
\end{equation*}

For the energy dissipation law, divide both sides by $U^2 L \eta$, and we obtain
$$
\begin{aligned}
    \mbox{Left} 
    & = \frac{\dd}{\dd \tilde{t}}  \int_{\tilde{\Omega}}    \frac{1}{2} {\rm Re} |\tilde{\uvec}|^2 + \frac{\epsilon_p}{{\rm Wi}} \bigg[ \int_{\mathbb{R}^d} \cdots \int_{\mathbb{R}^d}  \tilde{f} (\ln  \tilde{f}-C) + \alpha_1 \tilde{\Psi} \tilde{f}  \dd \tilde{\qvec}_1 \cdots \dd \tilde{\qvec}_{N-1} \bigg]  \ \dd \tilde{\x}
\end{aligned}
$$
with
$${\rm Re} = \frac{\rho U L}{\eta}, \quad {\rm Wi} = \frac{\lambda U  }{L}, \quad \tilde{\eta}_s = \frac{\eta_s}{\eta}, \quad \epsilon_p = \frac{\eta_p}{\eta}, \quad \lambda = \frac{\zeta}{4H}.$$
Here, $\eta_p = \lambda_p k_B T \lambda$ is related to the polymer viscosity, $\eta$ is the total fluid viscosity and $\eta = \eta_s + \eta_p$.  

$$
\begin{aligned}
    \mbox{Right} 
    & = -\int_{\tilde{\Omega}}  \tilde{\eta}_s |\tilde{\nabla} \tilde{\uvec}|^2 - 4 \alpha_1  \epsilon_p \int_{\mathbb{R}^d} \cdots  \int_{\mathbb{R}^d}  \tilde{f}   \sum_{i=1}^{N-1} \sum_{j=1}^{N-1} C_{ij} ( \dot{\tilde{\qvec}}_i - (\tilde{\nabla} \tilde{\uvec})^T \cdot  \tilde{\qvec}_i) \\
     & \cdot ( \dot{\tilde{\qvec}}_j - (\tilde{\nabla} \tilde{\uvec})^T \cdot  \tilde{\qvec}_j ) \dd \tilde{\qvec}_1 \cdots \dd \tilde{\qvec}_{N-1} \dd \tilde{\x}. \\
\end{aligned}
$$


Therefore, the final governing system is given by

\begin{equation}\label{CPsystemwithf_nondim}
\left\{
\begin{aligned}
&{\rm Re} (\uvec_t+\uvec \cdot \nabla_{\x} \uvec)+\nabla_{\x} p= (1-\epsilon_p) \Delta_{\x} \uvec+\nabla_{\x} \cdot {\bm \tau}_p,\\
& \nabla_{\x}\cdot\uvec=0,\\
&{\bm \tau}_p 
= \sum_{i=1}^{N-1}  \frac{\epsilon_p \alpha_1}{{\rm Wi}} \int_{\mathbb{R}^d} \cdots \int_{\mathbb{R}^d} f  \nabla_{\qvec_i} \Psi \otimes \qvec_i \dd \qvec_1 \cdots \dd \qvec_{N-1},\\
&f_t + \nabla_{\x} f \cdot \uvec + \sum_{i=1}^{N-1} \nabla_{\qvec_i} \cdot ( (\nabla_{\x} \uvec)^T \cdot  \qvec_i f) =  \frac{1}{4 \alpha_1 {\rm Wi}}  \sum_{i=1}^{N-1} \nabla_{\qvec_i} \cdot \bigg(  \sum_{k=1}^{N-1}   A_{ik}  [\nabla_{\qvec_k} f+ \alpha_1 f \nabla_{\qvec_k} \Psi] \bigg).
\end{aligned}
\right.
\end{equation}

\section{Numerical Method}

An efficient numerical method has been proposed in Refs. \cite{bao2025jcp, bao2024prf}, to solve the classical dumbbell (two-bead) micro-macro model, by incorporating a deterministic particle method for the macroscopic Fokker-Planck equation with a finite element discretization for the macroscopic fluid equation \cite{bao2021, becker2008,chenrui2015}. 
In Ref. \cite{bao2025jcp}, it was indicated that this deterministic-particle-FEM numerical method is an effective method which is capable of capturing the rich behaviors of polymeric fluids.
%
%
%

In this article, the deterministic-particle-FEM numerical method is employed for the micro-macro multi-bead model \eqref{CPsystemwithf_nondim}. 
In this section, we first present the deterministic particle approximation of $f$ and the final coarse-grained micro-macro particle system. 
Precisely, for the $N$-bead model, since
$$
 \int_{\mathbb{R}^d} \cdots  \int_{\mathbb{R}^d}  f (\x, \qvec_1, \qvec_2, \cdots, \qvec_{N-1}, t) \dd \qvec_1 \dd \qvec_2 \cdots \dd\qvec_{N-1} = 1,
$$
using the particle approximation at the microscopic level, for fixed $\x \in \Omega$and $t$, $f (\x, \qvec_1, \cdots, \qvec_{N-1}, t)$ can be approximated by 
\begin{equation}\label{fapproximation}
\begin{aligned}
 f (\x, \qvec_1, \qvec_2, \cdots, \qvec_{N-1}, t) & \approx f_{N_p} (\x, \qvec_1, \qvec_2, \cdots, \qvec_{N-1}, t) \\
 & = f_{N_p} (\x, \bar{\Qvec}, t) = \sum_{J=1}^{N_p}  \omega_J(\x, t)  \delta(\bar{\Qvec} - \bar{\Qvec}^J(\x, t)).
\end{aligned}
\end{equation}
%
Here, $N_p$ is the number of particles at $\x$ and time $t$. And $\bar{\Qvec}$, combined from $\qvec_1, \qvec_2,  \cdots, \qvec_{N-1}$, is a long vector of size $( (N-1)d, 1)$:
$$\bar{\Qvec} = (\qvec_1, \qvec_2,  \cdots, \qvec_{N-1})^T, \quad \bar{\Qvec}^J = (\qvec_{1}^J, \qvec_{2}^J,  \cdots, \qvec_{N-1}^J )^T.$$ 
And $\{\bar{\Qvec}^J(\x, t) \}_{J=1}^{N_p}$ is a set of representative particles at $\x$ at time $t$, $\omega_J(\x, t)$ is the weight of the corresponding particle satisfying $\sum_{J=1}^{N_p}\omega_J(\x, t) = 1$. In the current work, we fix $\omega_J(\x, t) = \frac{1}{N_p}$, i.e., all the particles are equally weighted.

The micro-macro multi-bead model with particles $\{\bar{\Qvec}^J(\x, t) \}_{J=1}^{N_p}$ can be obtained via utilizing the discrete energetic variational approach \cite{wang2020jcp, wangEVI, bao2025jcp}.
Via the "Approximation-then-Variation" approach, we first substitute the approximation \eqref{fapproximation} into the continuous energy dissipation law, then a discrete energy-dissipation law in terms of particles and the macroscopic flow can be obtained: 
$$
\frac{\dd}{\dd t} \mathcal{F}_h \left[ \{ \bar{\Qvec}^I\}_{I=1}^{N_p}, \x \right] = -2\mathcal{D}_h \left[ \{\bar{\Qvec}^I\}_{I=1}^{N_p}, \bigg\{\frac{D \bar{\Qvec}^I}{Dt}\bigg\}_{I=1}^{N_p}, \x, \uvec  \right]. 
$$
The discrete energy is given by
\begin{equation}
\begin{aligned}
    \mathcal{F}_h \left[ \{ \bar{\Qvec}^I\}_{I=1}^{N_p}, \x \right]& = \mathcal{F}_h^{macro} + \int_{\Omega} \mathcal{F}_h^{micro} \dd \x\\
    & = \int_{\Omega}  \frac{1}{2} {\rm Re} |\uvec|^2 \dd \x+ \int_{\Omega} \frac{\epsilon_p}{{\rm Wi}} \frac{1}{N_p} \sum_{I=1}^{N_p} \left[ \ln \left(\frac{1}{N_p} \sum_{J=1}^{N_p} K_h( \bar{\Qvec}^I -  \bar{\Qvec}^J )\right)+ \alpha_1 \Psi( \bar{\Qvec}^I)\right] \dd \x.
\end{aligned}
\end{equation}
Here, we introduce a kernel regularization for the term $\ln f_{N_p} (\x,  \bar{\Qvec}, t)$, namely, replacing $\ln f_{N_p}$ by $\ln (K_h\ast f_{N_p})$, where $K_h$ is a kernel function and
$$
K_h\ast f_{N_p}(:, \bar{\Qvec}, :)=\int K_h( \bar{\Qvec}-\hat{\Qvec}) f_{N_p}(:, \bar{\Qvec}, :) \dd \hat{\Qvec} =\frac{1}{N_p} \sum_{J=1}^{N_p} K_h(\bar{\Qvec}- \bar{\Qvec}^J(\x, t)).
$$
 A typical choice of $K_h$ is the Gaussian kernel, given by
$$
K_h(\Qvec_1, \Qvec_2)=\frac{1}{(\sqrt{2\pi}h_p)^{(N-1)\times d}}\exp\left(-\frac{|\Qvec_1-\Qvec_2|^2}{2h_p^2}\right).
$$
Here, $h_p$ is the bandwidth which controls the inter-particle
distances 
in the configuration space.


And the discrete dissipation term is given by
\begin{equation}
\begin{aligned}
& -2\mathcal{D}_h \left[ \{\bar{\Qvec}^I\}_{I=1}^{N_p}, \bigg\{\frac{D \bar{\Qvec}^I}{Dt}\bigg\}_{I=1}^{N_p}, \x, \uvec  \right] \\  
&\qquad = -2\mathcal{D}_h^{macro} - \int_{\Omega} 2 \mathcal{D}_h^{micro} \dd \x\\
&
\qquad = - \int_{\Omega} (1-\epsilon_p) |\nabla \uvec|^2 \dd \x- \int_{\Omega} 4 \alpha_1  \epsilon_p   \frac{1}{N_p}  \sum_{I=1}^{N_p} \sum_{i=1}^{N-1}  \sum_{j=1}^{N-1}  C_{ij}( \dot{\qvec}_i^I - (\nabla_{\x} \uvec)^T \cdot \qvec_i^I ) \cdot ( \dot{\qvec}_j^I - (\nabla_{\x} \uvec)^T \cdot \qvec_j^I )  \dd \x.  \\
\end{aligned}
\end{equation}

Within the microscopic discrete free energy $\mathcal{F}_h^{micro}$ and discrete dissipation $\mathcal{D}_h^{micro}$ on the microscopic scale, the dynamics of particles $\{\bar{\Qvec}^I(\x, t) \}_{J=1}^{N_p}$ can be derived by performing the discrete energetic variational approach via
$$
\frac{\delta \mathcal{D}_h^{micro}}{\delta \dot{\bar{\Qvec}}^I}=-\frac{\delta \mathcal{F}_h^{micro}}{\delta \bar{\Qvec}^I}. 
$$

By simple computation, we obtain
\begin{equation}
\frac{\delta \mathcal{F}_h^{micro}}{\delta \bar{\Qvec}^I} = 
\bm{\mu}^I,    
\end{equation}
where 
\begin{equation}\label{mudef}
    \bm{\mu}^I = \frac{\epsilon_p}{{\rm Wi}} \frac{1}{N_p} \bigg \{ \bigg(   \frac{\sum_{J=1}^{N_p} \nabla_{\bar{\Qvec}^I} K_h(\bar{\Qvec}^I-\bar{\Qvec}^J )}{ \sum_{J=1}^{N_p}  K_h( \bar{\Qvec}^I - \bar{\Qvec}^J ) }  +  \sum_{K=1}^{N_p}  \frac{ \nabla_{\bar{\Qvec}^I} K_h(\bar{\Qvec}^K-\bar{\Qvec}^I)}{ \sum_{J=1}^{N_p}K_h(\bar{\Qvec}^K-\bar{\Qvec}^J)} \bigg)   + \alpha_1 \nabla_{\bar{\Qvec}^I} \Psi (\bar{\Qvec}^I) \bigg\},
\end{equation}
and
\begin{equation}
\frac{\delta \mathcal{D}_h}{\delta \dot{\bar{\Qvec}}^I} = 2\alpha_1 \epsilon_p \frac{1}{N_p}  2 \bm{\mathcal{C}} ( \dot{\bar{\Qvec}}^I - (\nabla_{\x} \Uvec)^T \cdot  \bar{\Qvec}^I).
\end{equation}
Here, $((N-1)d) \times ((N-1)d)$ matrixes $\bm{\mathcal{C}}$ and $\nabla_{\x}\Uvec$ are given by 
$$
\bm{\mathcal{C}}=\begin{bmatrix}
C_{11} \bm I_d & C_{12} \bm I_d& \cdots & C_{1, N-1} \bm I_d\\
C_{21} \bm I_d&  C_{22} \bm I_d& \cdots & C_{2, N-1} \bm I_d\\
& & \ddots & & \\
C_{N-1, 1} \bm I_d& C_{N-1, 2} \bm I_d & \cdots & C_{N-1, N-1} \bm I_d\\
\end{bmatrix}_{((N-1)d) \times ((N-1)d)}
$$
with $C_{ij}$ the components of the Kramers matrix, 
$$
\nabla_{\x}\Uvec = 
\begin{bmatrix}
\nabla_{\x}\uvec & & & & \\
& \nabla_{\x}\uvec & &  &\\
& & \ddots & & \\
& & & & \nabla_{\x}\uvec
\end{bmatrix}_{((N-1)d) \times ((N-1) d)}.
$$
It leads to the particle equation
\begin{equation}\label{Eq_qi}
\begin{aligned}
& \bm{\mathcal{C}} ( \dot{\bar{\Qvec}}^I - (\nabla_{\x} \Uvec)^T \cdot  \bar{\Qvec}^I) \\
& \qquad = - \frac{1}{4 \alpha_1 {\rm Wi} } \bigg[ \bigg(   \frac{\sum_{J=1}^{N_p} \nabla_{\bar{\Qvec}^I} K_h(\bar{\Qvec}^I-\bar{\Qvec}^J )}{ \sum_{J=1}^{N_p}  K_h( \bar{\Qvec}^I - \bar{\Qvec}^J ) }  +  \sum_{K=1}^{N_p}  \frac{ \nabla_{\bar{\Qvec}^I} K_h(\bar{\Qvec}^K-\bar{\Qvec}^I)}{ \sum_{J=1}^{N_p}K_h(\bar{\Qvec}^K-\bar{\Qvec}^J)} \bigg)   + \alpha_1 \nabla_{\bar{\Qvec}^I} \Psi (\bar{\Qvec}^I) \bigg].
\end{aligned}
\end{equation}
It can also be written as 
\begin{equation}\label{Eq_qi_2}
\begin{aligned}
& \dot{\bar{\Qvec}}^I - (\nabla_{\x} \Uvec)^T \cdot  \bar{\Qvec}^I \\
& \qquad = - \frac{1}{4 \alpha_1 {\rm Wi} } \bm{\mathcal{A}} \bigg[ \bigg(   \frac{\sum_{J=1}^{N_p} \nabla_{\bar{\Qvec}^I} K_h(\bar{\Qvec}^I-\bar{\Qvec}^J )}{ \sum_{J=1}^{N_p}  K_h( \bar{\Qvec}^I - \bar{\Qvec}^J ) }  +  \sum_{K=1}^{N_p}  \frac{ \nabla_{\bar{\Qvec}^I} K_h(\bar{\Qvec}^K-\bar{\Qvec}^I)}{ \sum_{J=1}^{N_p}K_h(\bar{\Qvec}^K-\bar{\Qvec}^J)} \bigg)   + \alpha_1 \nabla_{\bar{\Qvec}^I} \Psi (\bar{\Qvec}^I) \bigg],
\end{aligned}
\end{equation}
where $((N-1)d) \times ((N-1)d)$ matrix $\bm{\mathcal{A}}$ is given by 
$$
\bm{\mathcal{A}}=\begin{bmatrix}
A_{11} \bm I_d & A_{12} \bm I_d& \cdots & A_{1, N-1} \bm I_d\\
A_{21} \bm I_d&  A_{22} \bm I_d& \cdots & A_{2, N-1} \bm I_d\\
& & \ddots & & \\
A_{N-1, 1} \bm I_d& A_{N-1, 2} \bm I_d & \cdots & A_{N-1, N-1} \bm I_d\\
\end{bmatrix}_{((N-1)d) \times ((N-1)d)}
$$
with $A_{ij}$ the components of the Rouse matrix.

The variational procedure for the macroscopic flow is almost the same as that in the continuous case, as shown in Refs. \cite{wang2021two, bao2025jcp}:
$$
\frac{\delta \mathcal{D}_h}{\delta \dot{\x}}=-\frac{\delta \mathcal{F}_h}{\delta \x}. 
$$
By simple computation, we obtain
$$
\frac{\delta \mathcal{F}_h}{\delta \x} =  {\rm Re} (\uvec_t + \uvec \cdot \nabla_{\x} \uvec)
$$
and
$$
\frac{\delta \mathcal{D}_h}{\delta \dot{\x}} = -(1-\epsilon_p) \Delta_{\x} \uvec + 4\alpha_1 \epsilon_p \frac{1}{N_p}  \sum_{I=1}^{N_p} \nabla_{\x} \cdot \bigg( 
 \sum_{i=1}^{N-1}  \sum_{j=1}^{N-1}  C_{ij} (\dot{\qvec}_i^I - (\nabla_{\x} \uvec)^T \cdot \qvec_i^I) \otimes  \qvec_j^I \bigg).
$$

And thus, we can define the stress as 
\begin{equation}
\begin{aligned}
{\bm \tau}_p &= -4\alpha_1 \epsilon_p \frac{1}{N_p}  \sum_{I=1}^{N_p} \bigg( 
 \sum_{i=1}^{N-1}  \sum_{j=1}^{N-1}  C_{ij} (\dot{\qvec}_i^I - (\nabla_{\x} \uvec)^T \cdot \qvec_i^I) \otimes  \qvec_j^I \bigg).
\end{aligned}
\end{equation}
Then substitute the definition of  $\bm{\mu}^I$ in Eq. \eqref{mudef}, we can rewrite the definition of ${\bm \tau}_p$ as follows, 
\begin{equation}\label{taupdef}
\begin{aligned}
{\bm \tau}_p &=  \frac{1}{N_p}  \sum_{I=1}^{N_p} \bigg( 
 \sum_{i=1}^{N-1}  \sum_{j=1}^{N-1}  C_{ij} \sum_{k=1}^{N-1}[A_{ik} \bm{\mu}_k^I]
 \otimes  \qvec_j^I \bigg),\\
\end{aligned}
\end{equation}
where $\bm{\mu}_k^I$ is a vector consistent of the $(k-1)\times d+1$ th to $ k\times d$ th components of $\bm{\mu}^I$.

And thus, the micro-macro particle system reads as follows, 
\begin{equation}\label{Eq_particle}
\left\{
\begin{aligned}
&{\rm Re} (\uvec_t+\uvec \cdot \nabla_{\x} \uvec)+\nabla_{\x} p= (1-\epsilon_p) \Delta_{\x} \uvec+\nabla_{\x} \cdot {\bm \tau}_p,\\
& \nabla_{\x}\cdot\uvec=0,\\
&{\bm \tau}_p =   \frac{1}{N_p}  \sum_{I=1}^{N_p} \bigg( 
 \sum_{i=1}^{N-1}  \sum_{j=1}^{N-1}  C_{ij} \sum_{k=1}^{N-1}[A_{ik} \bm{\mu}_k^I]
 \otimes  \qvec_j^I \bigg),\\
& \dot{\bar{\Qvec}}^I - (\nabla_{\x} \Uvec)^T \cdot  \bar{\Qvec}^I \\
& \qquad = - \frac{1}{4 \alpha_1 {\rm Wi} } \bm{\mathcal{A}} \bigg[ \bigg(   \frac{\sum_{J=1}^{N_p} \nabla_{\bar{\Qvec}^I} K_h(\bar{\Qvec}^I-\bar{\Qvec}^J )}{ \sum_{J=1}^{N_p}  K_h( \bar{\Qvec}^I - \bar{\Qvec}^J ) }  +  \sum_{K=1}^{N_p}  \frac{ \nabla_{\bar{\Qvec}^I} K_h(\bar{\Qvec}^K-\bar{\Qvec}^I)}{ \sum_{J=1}^{N_p}K_h(\bar{\Qvec}^K-\bar{\Qvec}^J)} \bigg)   + \alpha_1 \nabla_{\bar{\Qvec}^I} \Psi (\bar{\Qvec}^I) \bigg].\\
\end{aligned}
\right.
\end{equation}

\bibliographystyle{elsarticle-num} 
\bibliography{MM}



\end{document}